\documentclass[12pt,journal,onecolumn]{IEEEtran}
\usepackage[hang]{subfigure}
\usepackage{epsfig,rotating,setspace,latexsym,amsmath,epsf,amssymb,amsfonts,bm,theorem,epstopdf}
\usepackage{cite,authblk}
\usepackage{algorithm}
\usepackage[noend]{algpseudocode}
\usepackage{setspace}
\usepackage{color}
\usepackage{graphicx}
\usepackage{mathrsfs}
\ifCLASSOPTIONcompsoc
\usepackage[caption=false, font=normalsize, labelfont=sf, textfont=sf]{subfig}
\else
\usepackage[caption=false, font=footnotesize]{subfig}
\fi

\newtheorem{theorem}{Theorem}
\newtheorem{lemma}{Lemma}

\newenvironment{Proof}[1]{\medskip\par\noindent{\bf Proof:\,}\,#1}{{\mbox{\,$\blacksquare$}\par}}

\allowdisplaybreaks

\definecolor{green1}{rgb}{0.2,0.7,0.2}
\definecolor{brown}{rgb}{1,0.5,0.2}

\usepackage[colorinlistoftodos,textwidth=\marginparwidth]{todonotes}

\begin{document}
\title{Version Age of Information in\\Clustered Gossip Networks \thanks{This work was supported by NSF Grants CCF 17-13977 and ECCS 18-07348. This paper was presented in part \cite{Buyukates21c} at IEEE SPAWC, Lucca, Italy, September 2021.}}
    \author{Baturalp Buyukates \qquad Melih Bastopcu  \qquad Sennur Ulukus\\
	\normalsize Department of Electrical and Computer Engineering\\
	\normalsize University of Maryland, College Park, MD 20742\\
	\normalsize  \emph{baturalp@umd.edu} \qquad \emph{bastopcu@umd.edu}  \qquad \emph{ulukus@umd.edu}}

\maketitle
\vspace{-2cm}
\begin{abstract}
We consider a network consisting of a single source and $n$ receiver nodes that are grouped into equal-sized clusters. Each cluster corresponds to a distinct community such that nodes that belong to different communities cannot exchange information. We use dedicated cluster heads in each cluster to facilitate communication between the source and the nodes within that cluster. Inside clusters, nodes are connected to each other according to a given network topology. Based on the connectivity among the nodes, each node relays its current stored version of the source update to its neighboring nodes by \emph{local gossiping}. We use the \emph{version age} metric to assess information freshness at the receiver nodes. We consider disconnected, ring, and fully connected network topologies for each cluster. For each of these network topologies, we characterize the average version age at each node and find the average version age scaling as a function of the network size $n$. Our results indicate that per node average version age scalings of $O(\sqrt{n})$, $O(n^{\frac{1}{3}})$, and $O(\log n)$ are achievable in disconnected, ring, and fully connected cluster models, respectively. Next, we increase connectivity in the network and allow gossiping among the cluster heads to improve version age at the receiver nodes. With that, we show that when the cluster heads form a ring network among themselves, we obtain per node average version age scalings of $O(n^{\frac{1}{3}})$, $O(n^{\frac{1}{4}})$, and $O(\log n)$ in disconnected, ring, and fully connected cluster models, respectively. Next, focusing on a ring network topology in each cluster, we introduce hierarchy to the considered clustered gossip network model and show that when we employ two levels of hierarchy, we can achieve the same $O(n^{\frac{1}{4}})$ scaling without using dedicated cluster heads. We generalize this result for $h$ levels of hierarchy and show that per user average version age scaling of $O(n^{\frac{1}{2h}})$ is achievable in the case of a ring network in each cluster across all hierarchy levels. Finally, we find the version age-optimum cluster sizes as a function of the source, cluster head, and node update rates through numerical evaluations.
\end{abstract}
 
\section{Introduction}

The age of information metric has been introduced in \cite{Kaul12a} to measure information timeliness in real-time status updating systems and has a wide range of promising applications in emerging time-critical technologies including next-generation holographic communications, autonomous systems, and smart factories. Age of information metric studies lie at the intersection of information, communication, networking, and queueing theory fields \cite{Kosta_Survey, SunSurvey, YatesSurvey}. 

The original age metric measures the time passed since the most recent information at the monitor was generated at the source node. This age metric increases linearly in time in the absence of update deliveries at the monitor. When an update is received, the age value drops to the age of the received update. This evolution in time demonstrates the fundamental limitation of the original age metric, which is the assumption that the age at the monitor continues to increase as time passes irrespective of any changes at the source side in the underlying source process. That is, even if the source information does not change and the monitor has the most up-to-date information, as time passes, the original age metric deems monitor's knowledge about the source process \emph{stale}. This may not necessarily be the case in many applications, including content delivery services and surveillance systems. To overcome this inherent challenge, in the age of information literature, several variants of the original age metric have been proposed. A common feature of these recently proposed age variants is the fact that the age of the monitor stays the same until the information at the source changes even if no updates are received at the monitor. Among these are binary freshness metric \cite{cho03, kolobov19a, Bastopcu2021, Bastopcu20e, Kaswan21, Bastopcu21f}, age of synchronization \cite{Zhong18c}, and age of incorrect information \cite{Maatouk20b, Maatouk20a, Kam20a}.

Similar in spirit, recently, a new age metric called \emph{version age} has appeared in the literature \cite{Yates21, Eryilmaz21}. In the version age context, each update at the source is considered a version change so that the version age counts how many versions out-of-date the information at the monitor is, compared to the version at the source. Unlike the original age metric, the version age has discrete steps such that the version age of a monitor increases by one when the source generates a newer version, i.e., fresher information. In between version changes at the source, version age of the monitor stays the same indicating that the monitor still has the most recent information. A predecessor of version age has appeared in \cite{Bastopcu20c}, which considers timely tracking of Poisson counting processes by minimizing the count difference, i.e, version difference, between the process and its estimate.

Recently, reference \cite{Yates21} has used the version age metric to characterize timeliness in memoryless gossip networks composed of $n$ arbitrarily connected nodes. In \cite{Yates21}, the source sends information to the receiver nodes by implementing a Poisson updating mechanism, i.e., with exponential inter-update times at each receiver node. Similar Poisson updating schemes have been investigated in the age literature in the context of social networks \cite{Ioannidis09}, timely tracking \cite{Bastopcu20c, Bastopcu21b}, and timely cache updating \cite{Bastopcu2021, Bastopcu20e, Kaswan21}. In addition to source delivering updates to the receiver nodes, each node in \cite{Yates21} relays their stored version of the source information to their neighboring nodes. Also referred to as \emph{gossiping}, this additional information exchange among the nodes improves the age scaling at the nodes since each node can receive updates from its neighbors as well as from the source node. As a result of this gossiping, \cite{Yates21} shows that the average version age scales as $O(\sqrt{n})$ in a bi-directional ring network and as $O(\log n)$ in a fully connected network, where $n$ is the number of nodes; note that the average version age would scale as $O(n)$ without gossiping, i.e., if the network is disconnected.

There have been significant efforts in the age literature to characterize and improve the average age scaling in large networks considering the classical age metric with possibly many source-destination pairs. Recent works have achieved $O(1)$ scaling in multicast networks \cite{Zhong17a, Zhong18b, Buyukates18, Buyukates18b, Buyukates19} using a centralized transmission scheme administered by the source, and $O(\log n)$ scaling in distributed peer-to-peer communication networks \cite{Buyukates19b, Buyukates21b} using a hierarchical local cooperation scheme. 

\begin{figure}[t]
	\centerline{\includegraphics[width=0.8\columnwidth]{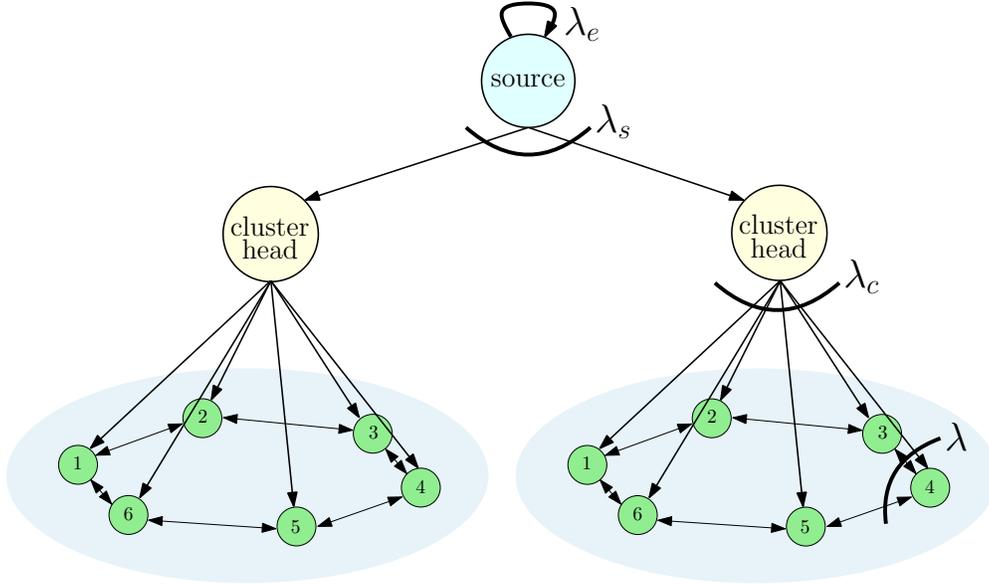}}
	\caption{Tiered network model where blue node at the center represents the source, yellow nodes represent the cluster heads, and green nodes represent the end users. Here, nodes in each cluster form a bi-directional ring network. Other possible network topologies within a cluster are shown in Fig.~\ref{Fig:netw_types}.}
	\label{Fig:disconnected}
\end{figure}

\begin{figure}[t]
	\centerline{\includegraphics[width=0.8\columnwidth]{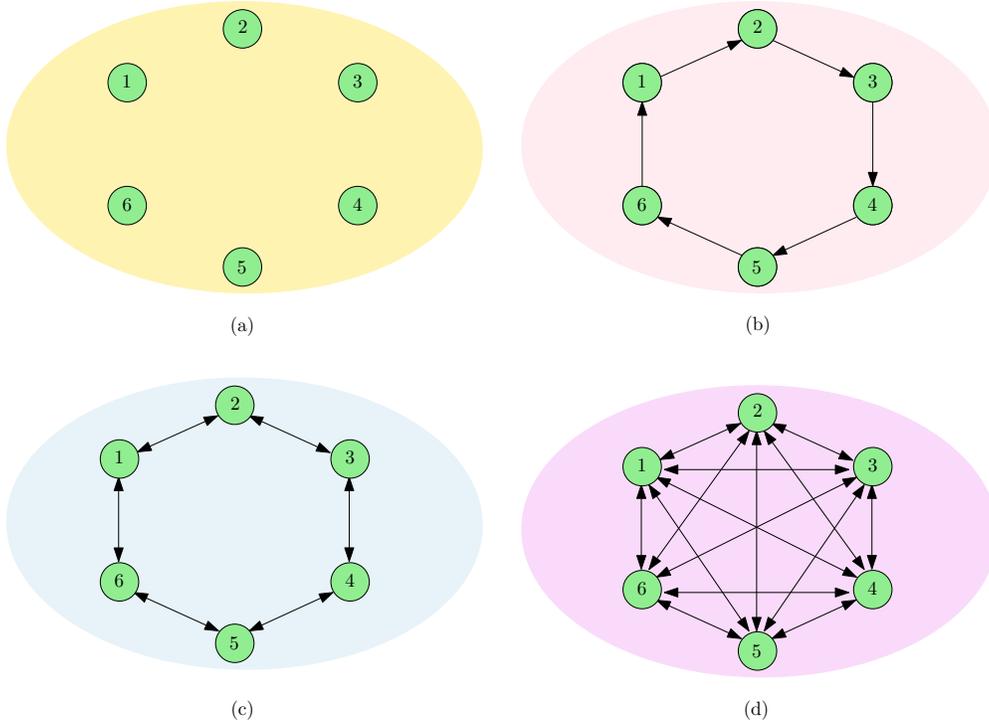}}
	\caption{Different network topologies that can be used within each cluster: (a) disconnected, (b) uni-directional ring, (c) bi-directional ring, and (d) fully connected. Fig.~\ref{Fig:disconnected} uses the one in (c).
	In this figure, cluster size is $k=6$.}
	\label{Fig:netw_types}
\end{figure}

Inspired by these, in this work, our aim is to study \emph{version age scaling} in more general gossip network models which exhibit a community structure; see Fig.~\ref{Fig:disconnected}. In our model, there is a single source node that generates updates following a Poisson process. Each such update constitutes a newer version of the underlying information process. The source updates multiple distinct communities regarding the underlying process. In our work, a community represents a set of receiver nodes clustered together which can only interact with each other. Each cluster has a dedicated cluster head, which serves that particular cluster. Akin to base stations in a cellular network, cluster heads act as gateways between the source and the receiver nodes in each cluster. Unlike the model in \cite{Yates21}, the source cannot directly deliver updates to receiver nodes in our model. Instead, source updates need to go through the corresponding cluster head to reach the receiver nodes in each cluster. There can be various degrees of gossip in each cluster, which we model by disconnected, uni-directional ring, bi-directional ring, and fully connected network topologies; see Fig.~\ref{Fig:netw_types}. Based on the underlying connectivity within clusters, we characterize the version age experienced by each node. In doing that, we employ the stochastic hybrid systems (SHS) approach \cite{Teel06, Hespanha07, Yates17a, Yates18b, Dogan20, Moltafet21} to develop recursive formulas that enable us to characterize the version age in arbitrarily connected clustered gossip networks. 

Additional hop constituted by the cluster heads between the source and the end-nodes presents us with opportunities to optimize the average version age scaling by carefully tuning the number of clusters and the cluster size. Specifically, our results indicate that even if the nodes within each community forego gossiping, i.e., disconnected networks within each cluster, we can achieve $O(\sqrt{n})$ scaling as opposed to $O(n)$. In addition, we obtain the same $O(\log n)$ scaling in the case of fully connected communities using fewer connections within clusters than \cite{Yates21}, and further reduce the scaling result in ring networks to $O(n^{\frac{1}{3}})$ from $O(\sqrt{n})$ in \cite{Yates21}. 

So far, the cluster heads do not participate in gossiping, i.e., cluster heads among themselves form a disconnected topology. To further improve the version age at the receiver nodes, next, we characterize the average version age and its scaling when the cluster heads form a ring network among themselves and exchange information. In that case, each cluster head uses some of its update rate to relay updates to its neighboring cluster heads while its remaining update rate is used to relay updates to the receiver nodes within its cluster. Thanks to the increased communication among the cluster heads, we can further improve the version age scaling to $O(n^{\frac{1}{3}})$ for disconnected networks within each cluster; to $O(n^{\frac{1}{4}})$ in the case of ring networks within each cluster. For the setup with a ring network in each cluster, we find the version age optimal update rate allocation at each cluster head. Interestingly, in the case of fully connected networks within each cluster, we find that the additional information exchange due to the gossip among the cluster heads does not improve the version age scaling. That is, the version age of an end user still scales as $O(\log n)$ even though cluster heads participate in gossip.

Motivated by the tiered structure in the clustered network model, next, we introduce hierarchy to our clustered network model. In this case, we forego cluster heads, and carefully place clusters of nodes in a hierarchical manner. That is, each node in a particular hierarchy level acts as a cluster head to a distinct cluster of nodes in the next hierarchy level. At the first level of hierarchy, we have a single cluster of $k_1$ nodes, each of which have a single cluster of $k_2$ nodes at the second level, and so on. Within the context of hierarchical clustered gossip networks, we consider a ring network in each cluster and show that the $O(\sqrt{n})$ scaling result of \cite{Yates21} and our cluster head-aided scaling result of $O(n^\frac{1}{3})$ for ring networks can be improved to $O(n^{\frac{1}{2h}})$ without the use of dedicated cluster heads, where $h$ denotes the number of hierarchy levels. For convenience, we provide the summary of all scaling results for version age in Table~\ref{table:results}. Finally, through numerical evaluations, we determine the version-age optimum cluster sizes for varying update rates employed by the source, cluster heads, and the nodes within each cluster.

\begin{table}[t]
\small
\centering
	\begin{center}
		\begin{tabular}{ | l | c | c | c |}
			\hline
			& $ \text{disconnected}$ & $\text{ring}$ & $\text{fully connected}$ \\ \hline
			$\text{no clustering as in \cite{Yates21}}$
			 & $O(n)$ & $O(\sqrt{n})$ & $O(\log n)$   \\ \hline
			$\text{clustered networks}$
			 & $O(\sqrt{n})$ & $O(n^{\frac{1}{3}})$ & $O(\log n)$   \\ \hline 
			 $\text{clustered networks with connected cluster heads}$
			 & $O(n^{\frac{1}{3}})$ & $O(n^{\frac{1}{4}})$ & $O(\log n)$   \\ \hline 
			 $\text{$h$-level hierarchical clustered networks }$
			 & $-$ & $O(n^{\frac{1}{2h}})$ & $-$   \\ \hline
		\end{tabular}
	\end{center}
	\caption{The summary of the scaling of version age in gossip networks. }
	\label{table:results}
	\vspace{-0.5cm}
\end{table}

The rest of this paper is organized as follows: In Section~\ref{sect:model}, we present our clustered gossip network model and the version age metric. In Section~\ref{sect:comm_age}, we characterize the average version age in clustered gossip networks considering disconnected, ring and fully connected network topologies in each cluster, and determine the corresponding version age scaling at a particular receiver node. In Section~\ref{sect:conn_ch}, we investigate the version age scaling improvement when the cluster heads are allowed to exchange information among themselves. Section~\ref{sect:hier} introduces the hierarchy concept into the considered clustered gossip network model to further improve the average version age performance and Section~\ref{sect:num_res} finds the optimal number of clusters and cluster sizes through numerical simulations as a function of the update rates at the source, cluster heads and nodes. Finally, we conclude this paper in Section~\ref{sect_discuss_conc} with a summary of the main results along with a discussion of some future directions. 

\section{System Model and the Age Metric}\label{sect:model}

We consider a system where a network of $n$ nodes is divided into $m$ clusters, each consisting of $k$ nodes such that $n = m k$ with $k, m \in \mathbb{Z}$; see Fig.~\ref{Fig:disconnected}. Each cluster is served by a distinct cluster head, which takes updates from the source and distributes them across that cluster. The source process is updated as a rate $\lambda_e$ Poisson process. The source has a total update injection rate of $\lambda_s$, which is uniformly distributed across cluster heads such that each cluster head is updated as a rate $\frac{\lambda_s}{m}$ Poisson process. From each cluster head to its corresponding cluster, the total update injection rate is $\lambda_{c}$ and this rate is uniformly allocated across the nodes in that cluster. That is, each node $i$ receives an update from its cluster head as a rate $\frac{\lambda_{c}}{k}$ Poisson process with $i \in \mathcal{N} \triangleq \{1, \ldots, n\}$.

Nodes in each cluster are connected to each other based on a connection graph. We consider varying levels of connectivity among nodes within each cluster. These are disconnected, uni-directional ring, bi-directional ring, and fully connected networks, which are shown in Fig.~\ref{Fig:netw_types} for a cluster of $k=6$ nodes. Updates received from the cluster head associated with each cluster are distributed across that cluster by utilizing the connections between the nodes. A node $i$ updates another node $j$ as a rate $\lambda_{ij}$ Poisson process. Each node in this system has a total update rate of $\lambda$, which is uniformly allocated to its neighboring nodes. That is, in the uni-directional ring, each node updates its neighbor node as a rate $\lambda$ Poisson process, whereas in bi-directional ring, each node has two neighboring nodes, each of which is updated as a rate $\frac{\lambda}{2}$ Poisson process. In the fully connected cluster, each node has $k-1$ neighbors each of which is updated as a rate $\frac{\lambda}{k-1}$ Poisson process. As a result of these local connections within a cluster, a node can receive different versions of the source update from its neighboring nodes in addition to the source updates received via its cluster head. 

To model the age at each node, we use the version age metric \cite{Yates21, Eryilmaz21}. We denote the version of the update at the source as $N_s(t)$, at cluster head $c$ as $N_{c}(t)$, with $c \in \mathcal{C} \triangleq \{1, \ldots, m\}$, and at node $i$ as $N_i(t)$, with $i \in \mathcal{N}$, at time $t$. The version age at node $i$ is given by $\Delta_i(t) = N_s(t)- N_i(t)$. Similarly, the version age at cluster head $c$ is  $\Delta_{c}(t) = N_{s}(t)- N_c(t)$. When node $i$ has the same version as the source, its version age becomes zero, i.e., $\Delta_i(t) = 0$. When the information at the source is updated, version ages at the cluster heads and the nodes increase by 1, e.g., $\Delta'_{c}(t) =\Delta_{c}(t)+1$. Each node $i$ can get updates either from its cluster head or the other nodes that it is connected to within its cluster. When node $i$ gets an update from its cluster head, its version age becomes 
\begin{align}
    \Delta'_i(t) = \min\{\Delta_{c}(t),\Delta_i(t)\} = \Delta_{c}(t). \label{clusterhead_update}
\end{align}
Last equality in (\ref{clusterhead_update}) follows since nodes in a cluster receive source updates through their cluster head so that they have either the same version or older versions of the information compared to their cluster head. When node $i$ receives an update from node $j$, its version age becomes
\begin{align}
    \Delta'_i(t) = \min\{\Delta_{i}(t),\Delta_j(t)\}. \label{node_update}
\end{align}
That is, node $i$'s version age is updated only if node $j$ has a newer version of the source information. Otherwise, the version age at node $i$ is not updated. 

\section{Version Age with Community Structure}\label{sect:comm_age}

In this section, we characterize the limiting version age of each node $i$, denoted by 
\begin{align}
  \Delta_i = \lim_{t \to \infty} \mathbb{E}[\Delta_i(t)], \quad i \in \{1, \ldots, n\},
\end{align}
considering various network topologies for the clusters. Since the network model in each cluster is identical and within each cluster the network is symmetric for each of the network topologies, age processes $\Delta_i(t)$ of all users are statistically identical. Thus, in the ensuing analysis, we focus on a single cluster $c \in \mathcal{C}$ and find the average version age of a node from that cluster. For this, we follow the construction in \cite{Yates21} and express ${\Delta}_i$ in terms of ${\Delta}_S$, which denotes the average version age of an arbitrary subset $S$ of the nodes in cluster $c$, where
\begin{align}
   {\Delta}_S(t) \triangleq \min_{j \in S} \Delta_j (t). 
\end{align}

We recall the following definitions from \cite{Yates21}: $\lambda_i(S)$ denotes the total update rate at which a node $i$ from cluster $c$ updates the nodes in set $S$. We have 
\begin{align}
    \lambda_i (S) = \begin{cases} 
     \sum_{j \in S} \lambda_{ij}, & i \notin S \\
     0, & i \in S.
   \end{cases}
\end{align} 
Similarly, $\lambda_{c}(S)$ denotes the total update rate of the cluster head of a particular cluster into the set $S$. Finally, set of updating neighbors of a set $S$ in cluster $c$ is 
\begin{align}
    N_c(S) = \{ i \in \mathcal{N}=\{1,\ldots,n \} : \lambda_i (S) > 0 \}. \label{neighbor_defn}
\end{align}
That is, the set $N_c(S)$ includes all updating neighbors of set $S$ in cluster $c$ excluding the cluster head. The total set of updating neighbors of set $S$ is given by $N(S) = c \cup N_c(S)$.

With these definitions, next, in Theorem~\ref{thm_age_set} below we give the resulting version age in our clustered system model as a specialization of \cite[Theorem~1]{Yates21}.

\begin{theorem}\label{thm_age_set}
 When the total network of $n$ nodes is divided into $m$ clusters, each of which consisting of a single cluster head and $k$ nodes with $n = m k$, the average version age of subset $S$ that is composed of nodes within a cluster $c$ is given by
 \begin{align}
     {\Delta}_S = \frac{\lambda_e + \lambda_{c}(S){\Delta}_{c} + \sum_{i \in N_c(S)} \lambda_i(S) {\Delta}_{S\cup \{i\}} }{\lambda_{c}(S) + \sum_{i \in N_c(S)} \lambda_i(S)},\label{Yates_recursion}
 \end{align}
 with ${\Delta}_{c} = m\frac{\lambda_e}{\lambda_s}$. 
\end{theorem}

\begin{Proof}
Proof of Theorem~\ref{thm_age_set} follows by applying \cite[Theorem~1]{Yates21} to our clustered network model and noting that updates arrive at the nodes through designated cluster heads. For completeness, we show the key steps in the proof below.

In our system, whenever there is an update being forwarded, a state transition occurs. We first present possible state transitions. We use $\mathcal{L}$ to denote the set of possible state transitions. Then, we have 
\begin{align}\label{transitions}
    \mathcal{L} = \{ (s,s)\} \cup \{ (s,c): c\in \mathcal{C}\} \cup \{(c,i): c \in \mathcal{C}, i \in \mathcal{N}\} \cup \{(i,j): i,j \in \mathcal{N} \},
\end{align}
where the first transition occurs when the source generates a new update, the second set of transitions occur when the source node updates a cluster head $c \in \mathcal{C}$. The third set of transitions occur when a cluster head $c$ updates a node in its cluster and finally the last set of transitions occur when an end user updates another end user from its cluster. In clustered gossip networks, different than \cite{Yates21}, as a result of transition $(i,j)$, the version age of an end user evolves as
\begin{align}\label{age_transitions}
    \Delta'_k = \begin{cases} 
      \Delta_k + 1, & i=j=s, k\in\mathcal{N}, \\
      \Delta_c, & i=c, j=k\in\mathcal{N}, \\
      \min(\Delta_i, \Delta_j), & i\in\mathcal{N}, j=k\in\mathcal{N},\\
      \Delta_k, & \text{otherwise},
   \end{cases} 
\end{align}
where $\Delta'_k$ is the version age of node $k$ after the transition. In (\ref{age_transitions}), the version age of node $k$ increases by one when the source generates a new update and becomes equal to the version age of its cluster head when node $k$ receives an update from its cluster head as explained in (\ref{clusterhead_update}). When node $k$ receives an update from another node in its cluster, its version age is updated only if the updating node has a newer version of the source information as shown in (\ref{node_update}).

Considering an arbitrary subset $S$ of nodes within a cluster with the version age evolution described in (\ref{age_transitions}), we deduce that after the $(s,s)$ transition, the version age of set $S$ is increased by one. For all other transitions $(i,j)$ with $j \in S$, we have
\begin{align}\label{versionage_update}
    \Delta'_S = \min_{k\in S} \Delta'_k = \min_{k \in S\cup\{i\}}\Delta_k = \Delta_{S\cup \{i\}}.
\end{align}
When $i=c$, from (\ref{versionage_update}), we have $\Delta'_S = \min_{k \in S\cup\{i\}}\Delta_k = \Delta_c$. If $j\notin S$, the version age of set $S$ is affected by transition $(i,j)$, i.e., $\Delta'_S = \Delta_S$. Using (\ref{age_transitions}) and (\ref{versionage_update}) and following similar steps as in \cite{Yates21} yields the result.
\end{Proof}

\subsection{Version Age in Clustered Disconnected Networks}\label{Sect:disconn}

Nodes in a cluster are not connected to each other. Thus, the network is a two-hop multicast network, where the first hop is from the source to $m$ cluster heads, and the second hop is from each cluster head to $k$ nodes; combine Fig.~\ref{Fig:disconnected} with Fig.~\ref{Fig:netw_types}(a). Multihop networks have been studied in \cite{Zhong17a, Zhong18b, Buyukates18, Buyukates18b, Buyukates19} considering the classical age metric, where the source keeps sending update packets until they are received by a certain number of nodes at each hop. We do not consider such centralized management of updates, but let the source update the cluster heads as Poisson processes, and let cluster heads forward these packets to the nodes within their clusters as further Poisson processes.

Let $S_1$ denote an arbitrary $1$-node subset of a cluster. Subset $S_1$ is only connected to the cluster head, i.e., $N_c(S_1) = \emptyset$. Using the recursion given in (\ref{Yates_recursion}), we find
\begin{align}
    {\Delta}_{S_1} = {\Delta}_{c} + k\frac{\lambda_e}{\lambda_{c}} = m\frac{\lambda_e}{\lambda_s} + k\frac{\lambda_e}{\lambda_{c}},\label{version_age_disconnect}
\end{align}
where ${\Delta}_{S_1}$ denotes the version age of a single node from the cluster. When the network consists of two-hops, version age is additive, in that the first term in (\ref{version_age_disconnect}) corresponds to the first hop and is equal to the version age at the cluster head, whereas the second term in (\ref{version_age_disconnect}) corresponds to the version age at the second hop between the cluster head and a node.

\begin{theorem}\label{corr_disconn}
  In a clustered network of disconnected users, the version age of a single user scales as $O(\sqrt{n})$.
\end{theorem}

Theorem~\ref{corr_disconn} follows by selecting $k=\sqrt{n}$ with $m = \frac{n}{k}=\sqrt{n}$ in (\ref{version_age_disconnect}) for fixed $\lambda_e$, $\lambda_s$, $\lambda_c$, which do not depend on $n$. Theorem~\ref{corr_disconn} indicates that when nodes are grouped into $\sqrt{n}$ clusters, an age scaling of $O(\sqrt{n})$ is achievable even though users forego gossiping. With the absence of cluster heads, i.e., when the source is uniformly connected to each of the $n$ users, the version age scaling of each disconnected user would be $O(n)$. By utilizing clusters, we incur an additional hop, but significantly improve the scaling result from $O(n)$ to $O(\sqrt{n})$. 
 
\subsection{Version Age in Clustered Ring Networks} \label{Sect:ring}

Nodes in each cluster form a ring network. We consider two types of ring clusters: uni-directional ring as shown in Fig.~\ref{Fig:netw_types}(b) and bi-directional ring as shown in Fig.~\ref{Fig:netw_types}(c).

First, we consider the uni-directional ring and observe that an arbitrary subset of $j$ adjacent nodes $S_j$ has a single neighbor node that sends updates with rate $\lambda$ for $j \leq k-1$. Each such subset $S_j$ receives updates from the cluster head with a total rate of $j\frac{\lambda_{c}}{k}$. Next, we use the recursion in (\ref{Yates_recursion}) to write
\begin{align}\label{version_age_ring_with_base}
  {\Delta}_{S_j} = \frac{\lambda_e+j\frac{\lambda_{c}}{k}{\Delta}_{c}+\lambda {\Delta}_{S_{j+1}}}{j\frac{\lambda_{c}}{k}+\lambda},  
\end{align}
for $j \leq k-1$ where ${\Delta}_{c}$ is the version age at the cluster head. We note that when $j=k$ the network becomes a simple two-hop network similar to that of Section~\ref{Sect:disconn} and we find ${\Delta}_{S_k} =m\frac{\lambda_e}{\lambda_s}+\frac{\lambda_e}{\lambda_{c}}$.

Next, we consider the bi-directional ring and observe that an arbitrary subset $S_j$ that consists of any adjacent $j$ nodes has two neighbor nodes, each with an incoming update rate of $\frac{\lambda}{2}$ for $j < k-1$. When $j=k-1$, $S_j$ has a single neighboring node that sends updates with a total rate $2\frac{\lambda}{2}=\lambda$. For $j\leq k-1$, the cluster head sends updates to subset $S_j$ with a total rate of $j\frac{\lambda_{c}}{k}$. With all these, when we apply the recursion in (\ref{Yates_recursion}), we obtain exactly the same formula given in (\ref{version_age_ring_with_base}). 
\begin{lemma}\label{corr_ring}
 Both uni-directional and bi-directional ring cluster models yield the same version age for a single node when each node in a cluster has a total update rate of $\lambda$.
\end{lemma}

Lemma~\ref{corr_ring} follows from the fact that either type of ring cluster induces the same recursion for an arbitrary subset of any adjacent $j$ nodes within a cluster as long as the total update rate per node $\lambda$ is the same. Thus, in the remainder of this paper, we only consider the bi-directional ring cluster model.

Before focusing on age scaling in a clustered network with a ring topology in each cluster, we revisit the ring network in \cite{Yates21}, and provide a proof of the $1.25 \sqrt{n}$ age scaling result observed therein as a numerical result. We show that the approximate theoretical coefficient is $\sqrt{\frac{\pi}{2}}=1.2533$. 

\begin{lemma}\label{yates_ring_proof}
  For the ring network model considered in \cite{Yates21}, the version age of a user scales as $\Delta_{S_1} \approx \sqrt{\frac{\pi}{2}}\frac{\lambda_e}{\lambda}\sqrt{n}$.
\end{lemma}

\begin{Proof}
From recursive application of \cite[Eqn.~(17)]{Yates21}, we obtain 
\begin{align}\label{yates_age_scaling_exp} 
    \Delta_{S_1} = \frac{\lambda_e}{\lambda}  \left(\sum_{i=1}^{n-1} a^{(n)}_i+a^{(n)}_{n-1}\right),
\end{align}
where $a^{(n)}_i$ is given for $i = 1,\ldots, n-1$ as
\begin{align} \label{defn_a_i}
a^{(n)}_i = \prod_{j=1}^{i}  \frac{1}{1+\frac{j}{n}}.
\end{align}
We note that $a^{(n)}_i$ decays fast in $i$, and consider $i=o(n)$,
\begin{align}
\!\!\!\!\!-\log(a^{(n)}_i) = \sum_{j=1}^{i} \log \left(1+\frac{j}{n}\right) \approx  \sum_{j=1}^{i} \frac{j}{n} = \frac{i (i+1)}{2n} \approx \frac{i^2}{n} \label{yates_age_scaling_exp2} \!\!
\end{align}
where we used $\log(1+x)\approx x$ for small $x$, and ignored the $i$ term relative to $i^2$. Thus, for small $i$, we have $a^{(n)}_i \approx e^{-\frac{i^2}{2n}}$. For large $i$, $a^{(n)}_i$ converges quickly to zero due to multiplicative terms in $\prod_{j=1}^{i} \frac{1}{1+j/n}$, and this approximation still holds. Thus, we have $\sum_{i=1}^{n-1} a^{(n)}_i \approx \sum_{i=1}^{n-1} e^{-\frac{i^2}{2n}}$. For large $n$, by using Riemann sum approximation with steps $\frac{1}{\sqrt{n}}$, we obtain  
\begin{align}\label{integral}
    \frac{1}{\sqrt{n}}\sum_{i=1}^{n-1} a^{(n)}_i \approx \frac{1}{\sqrt{n}}\sum_{i=1}^{n-1} e^{-\frac{i^2}{2n}} = \int_{0}^{\infty} e^{-\frac{t^2}{2}} \,dt=\sqrt{\frac{\pi}{2}}. 
\end{align}
Thus, we get $\sum_{i=1}^{n-1} a^{(n)}_i \approx\sqrt{\frac{\pi}{2}}\sqrt{n} $. By inserting this in (\ref{yates_age_scaling_exp}), we obtain the age scaling of a user as $\Delta_{S_1} \approx   \sqrt{\frac{\pi}{2}}\frac{\lambda_e}{\lambda}\sqrt{n}$.
\end{Proof}

Next, we focus on age scaling in a clustered network with a ring topology in each cluster. 
From recursive application of (\ref{version_age_ring_with_base}) along with ${\Delta}_{S_k}$, we obtain
\begin{align}\label{scaling_exact}
    \!\!\!\!\Delta_{S_1} =&\frac{\lambda_e}{\lambda}\left(\sum_{i=1}^{k-1} b^{(k)}_i \right)+\Delta_c\left(1-b^{(k)}_{k-1}\right)
    +\Delta_{S_k} b^{(k)}_{k-1},
\end{align}
where similar to (\ref{defn_a_i}), $b^{(k)}_i$ is given for $i=1,\ldots, k-1$ as
\begin{align} \label{defn_b_i}
b^{(k)}_i = \prod_{j=1}^{i} \frac{1}{1+\frac{j}{k}\frac{\lambda_c}{\lambda}}. 
\end{align}
When $k$ is large, $b^{(k)}_{k-1}$ goes to zero, and $\Delta_{S_1}$ in (\ref{scaling_exact}) becomes 
\begin{align} \label{ring_approximated_age_final}
    \Delta_{S_1}\approx \frac{\lambda_e}{\lambda} \left(\sum_{i=1}^{k-1} b^{(k)}_i \right)+\Delta_c \!\approx   \sqrt{\frac{\pi}{2}}\frac{\lambda_e}{\sqrt{\lambda\lambda_c}}\sqrt{k}+m\frac{\lambda_e}{\lambda_s},
\end{align}
where the second approximation follows as in the proof of Lemma~\ref{yates_ring_proof}. Terms in (\ref{ring_approximated_age_final}) are $O(\sqrt{k})$ and $O(m)$, respectively. In \cite{Yates21}, there is a single cluster, i.e., $m=1$ and $k=n$, and thus, the version age scaling is $O(\sqrt{n})$. In our model, by carefully adjusting the number of clusters and the cluster sizes, we can improve this $O(\sqrt{n})$ scaling result to $O(n^{\frac{1}{3}})$.
 
\begin{theorem}\label{ring_scaling}
  In a clustered network with a ring topology in each cluster, the version age of a single user scales as $O(n^{\frac{1}{3}})$.
\end{theorem}

Theorem~\ref{ring_scaling} follows by selecting $m=n^{\frac{1}{3}}$ with $k=\frac{n}{m}=n^{\frac{2}{3}}$ in (\ref{ring_approximated_age_final}) for fixed $\lambda_e$, $\lambda_s$, $\lambda_c$, $\lambda$, which do not depend on $n$.

\subsection{Version Age in Clustered Fully Connected Networks}\label{Sect:fully}

Nodes in each cluster form a fully connected network where each node is connected to all the other nodes within its cluster with rate $\frac{\lambda}{k-1}$. We find the version age for a subset of $j$ nodes $S_j$ in a cluster. Each such subset $j$ has $k-j$ neighbor nodes in addition to the cluster head associated with their cluster. Using the recursion given in (\ref{Yates_recursion}), we find
\begin{align}\label{version_age_fully_with_base}
  {\Delta}_{S_j} = \frac{\lambda_e+{\frac{j\lambda_{c}}{k}{\Delta}_{c}}+\frac{j(k-j)\lambda}{k-1} {\Delta}_{S_{j+1}} }{\frac{j\lambda_{c}}{k}+\frac{j(k-j)\lambda}{k-1}},  
\end{align}
for $j \leq k-1$, where ${\Delta}_{c}$ is equal to $m\frac{\lambda_e}{\lambda_s}$. The average version age of the whole cluster is ${\Delta}_{S_k} = \Delta_c + \frac{\lambda_e}{\lambda_c} = m \frac{ \lambda_e}{\lambda_s}+\frac{\lambda_e}{\lambda_{c}}$. 

Next, we present bounds for ${\Delta}_{S_1}$.

\begin{lemma}\label{lemma_fully}  
When $\lambda_{c}=\lambda$, in a clustered network with fully connected topology in each cluster, the version age of a single node satisfies
 \begin{align}\label{version_age_fully_with_base_simp_6}
   \frac{(k-1)^2+k}{k^2} \Delta_c+ \frac{\lambda_e}{\lambda}\left(\frac{k-1}{k}\sum_{\ell=1}^{k-1}\frac{1}{\ell}+ \frac{1}{k} \right) \leq {\Delta}_{S_1} \leq \Delta_c + \frac{\lambda_e}{\lambda} \left(\sum_{\ell=1}^{k}\frac{1}{\ell} \right). 
 \end{align}
\end{lemma}

\begin{Proof}
We use steps similar to those in the proof of \cite[Theorem~2]{Yates21} and also consider the additional hop from the source to the cluster heads. For $\lambda_{c} = \lambda$, we take $j=k-\ell$ and (\ref{version_age_fully_with_base}) becomes
\begin{align}\label{version_age_fully_with_base_simp_1}
  {\Delta}_{S_{k-\ell}} = \frac{\frac{1}{k-\ell}\frac{\lambda_e}{\lambda} + \frac{1}{k} \Delta_c + \frac{\ell}{k-1}{\Delta}_{S_{k-\ell+1}} }{\frac{1}{k}+\frac{\ell}{k-1}},
\end{align}
for $\ell \leq k-1$ and ${\Delta}_{S_k} = \Delta_c +\frac{\lambda_e}{\lambda}$, where $\Delta_c$ is the age at the cluster head. Defining $\hat{\Delta}_{S_\ell} \triangleq {\Delta}_{S_{k-\ell+1}}$, we get
\begin{align}\label{version_age_fully_with_base_simp_2}
  \hat{\Delta}_{S_{\ell+1}} = \frac{ \frac{1}{k-\ell}\frac{\lambda_e}{\lambda} + \frac{1}{k}\Delta_c +  \frac{\ell}{k-1}\hat{\Delta}_{S_\ell} }{\frac{1}{k}+\frac{\ell}{k-1}}.  
\end{align}
Next, one can show that $\hat{\Delta}_{S_{\ell+1}}$ satisfies the following
\begin{align}\label{version_age_fully_with_base_simp_3}
    \hat{\Delta}_{S_{\ell+1}} \leq \frac{ \frac{1}{k-\ell}\frac{\lambda_e}{\lambda} + \frac{1}{k}\Delta_c +
    \frac{\ell}{k}\hat{\Delta}_{S_{\ell}} }{\frac{1}{k}+\frac{\ell}{k}}.
\end{align}
Defining $\tilde{\Delta}_{S_\ell} \triangleq \frac{\ell}{k}\hat{\Delta}_{S_{\ell}}$ and plugging it in (\ref{version_age_fully_with_base_simp_3}), we get
\begin{align}\label{version_age_fully_with_base_simp_4}
    \tilde{\Delta}_{S_{\ell+1}} = \frac{\ell+1}{k}\hat{\Delta}_{S_{\ell}} \leq \frac{1}{k-\ell}\frac{\lambda_e}{\lambda} + \frac{1}{k}\Delta_c + \tilde{\Delta}_{S_{\ell}}.
\end{align}
Noting that $\tilde{\Delta}_{S_1} = \frac{\hat{\Delta}_{S_1}}{k} = \frac{{\Delta}_{S_k}}{k} = \frac{1}{k}\left(\Delta_c +\frac{\lambda_e}{\lambda}\right)$, we write
\begin{align}\label{version_age_fully_with_base_simp_5}
    \tilde{\Delta}_{S_k} \leq \Delta_c + \frac{\lambda_e}{\lambda} \left(\sum_{\ell=1}^{k}\frac{1}{\ell} \right).
\end{align}
Since $\tilde{\Delta}_{S_k} = \hat{\Delta}_{S_k} = {\Delta}_{S_1}$,  (\ref{version_age_fully_with_base_simp_5}) presents an upper bound to the version age of a single node. For the lower bound, we start with (\ref{version_age_fully_with_base_simp_2}) and observe that we have
\begin{align}
    \hat{\Delta}_{S_{\ell+1}}  \geq \frac{k-1}{\ell+1} \left(\frac{1}{k-\ell}\frac{\lambda_e}{\lambda} + \frac{1}{k}\Delta_c +  \frac{\ell}{k-1}\hat{\Delta}_{S_\ell}\right). \label{lowerbound_1}
\end{align} 
Defining $\bar{\Delta}_{S_{\ell}} \triangleq \frac{\ell}{k-1}\hat{\Delta}_{S_{\ell}}$ and using it in (\ref{lowerbound_1}) gives
\begin{align}
    \bar{\Delta}_{S_{\ell+1}} = \frac{\ell+1}{k-1}\hat{\Delta}_{S_{\ell+1}}  \geq \frac{1}{k-\ell}\frac{\lambda_e}{\lambda} + \frac{1}{k}\Delta_c +  \bar{\Delta}_{S_\ell}. \label{lowerbound_2}
\end{align}
Starting with the fact that $\bar{\Delta}_{S_1} = \frac{1}{k-1}\hat{\Delta}_{S_1} = \frac{m+1}{k-1} \frac{\lambda_e}{\lambda}$, the recursion in (\ref{lowerbound_2}) yields
\begin{align}
    \Delta_{S_1} \geq \frac{(k-1)^2+k}{k^2} \Delta_c+ \frac{\lambda_e}{\lambda}\left(\frac{k-1}{k}\sum_{\ell=1}^{k-1}\frac{1}{\ell}+ \frac{1}{k} \right),
\end{align}
upon noting that $\Delta_{S_1} = \frac{k-1}{k} \bar{\Delta}_{S_k}$, which concludes the proof of the lemma.
\end{Proof}

From (\ref{version_age_fully_with_base_simp_6}), we see that for large $n$ with $\lambda_c = \lambda$, the version age of a single node $\Delta_{S_1}$ satisfies
\begin{align}\label{approximation}
\Delta_{S_1} \approx m\frac{\lambda_e}{\lambda_s} + \frac{\lambda_e}{\lambda}\log k.
\end{align}

\begin{theorem}\label{fully_conn_scaling}
  In a clustered network with a fully connected topology in each cluster, the version age of a single user scales as $O(\log{n})$.
\end{theorem}

Theorem~\ref{fully_conn_scaling} follows in multiple different ways. For instance, it follows by selecting $m=1$ and $k =\frac{n}{m}=n$. That is, we have a single fully connected network of $n$ users as in \cite{Yates21}. Theorem~\ref{fully_conn_scaling} also follows by selecting $m=\log n$ and $k=\frac{n}{m}=\frac{n}{\log n}$. That is, we have $\log(n)$ fully connected clusters with $\frac{n}{\log n}$ users in each cluster. Thus, version age obtained under a smaller cluster size with less connections is the same as that obtained when all nodes are connected to each other. In particular, in our model with $m=\log n$, each node has $O(\frac{n}{\log n})$ connections in comparison to $O(n)$ in \cite{Yates21}.

Finally, we note that, a recurring theme in the analysis of clustered networks is the fact that the version age at an end-node $\Delta_{S_1}$ is \emph{almost} additive in the version age at the cluster head $\Delta_c$ as seen in (\ref{version_age_disconnect}), (\ref{scaling_exact}), and (\ref{version_age_fully_with_base_simp_6}). It is \emph{exactly} additive in the case of disconnected clusters in (\ref{version_age_disconnect}).   

\section{Version Age with Community Structure Under Connected Cluster Heads}\label{sect:conn_ch}

\begin{figure}[t]
	\centerline{\includegraphics[width=0.5\columnwidth]{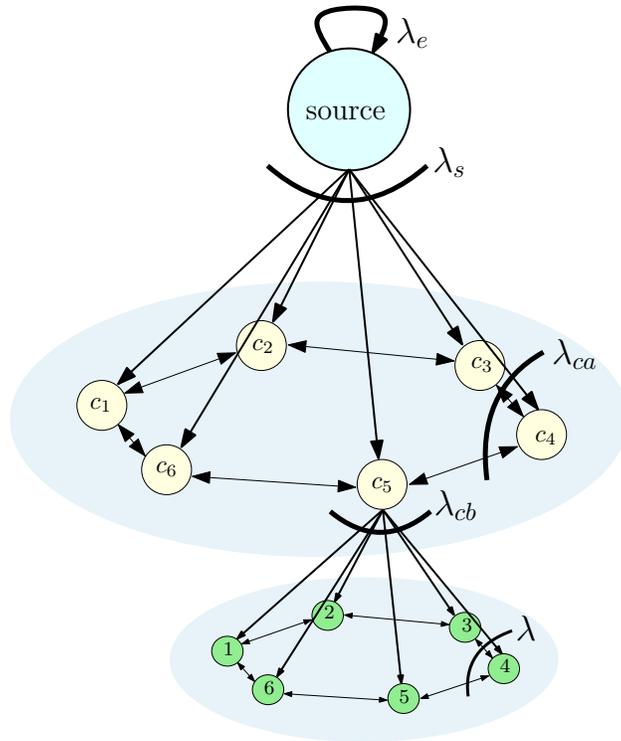}}
	\caption{Tiered network model where blue node represents the source, yellow nodes represent the cluster heads $c_1, \ldots, c_6$, and green nodes represent the end users. Here, cluster heads form a bi-directional ring network with $m=6$. Each cluster is associated with a cluster of $k=6$ nodes. Here, only one such cluster is shown. Nodes in each cluster form a bi-directional ring network. Other possible network topologies within a cluster are shown in Fig.~\ref{Fig:netw_types}.}
	\label{Fig:ring_clusterheads}
\end{figure}

So far, we have studied the cases in which the cluster heads are disconnected among themselves, and consequently, they do not exchange information with each other. In this section, we model the connectivity among the cluster heads with a bi-directional ring (see Fig.~\ref{Fig:ring_clusterheads}).\footnote{The model studied in Section~\ref{sect:comm_age} corresponds to the case in which the cluster heads form a disconnected topology.} \footnote{In addition to the bi-directional ring topology, one can study the version age considering fully connected cluster heads, which is omitted here to keep the discussion focused.} Thus, in this section, at the first tier, we have a ring network of $m$ cluster heads, each of which is serving its own cluster. Nodes in each cluster form a disconnected, ring, or fully connected network. Our aim in this section, is to analyze the effect of additional information exchange among the cluster heads on the average version age experienced by the end nodes.

When there is no information exchange among the cluster heads, i.e., disconnected cluster heads, each cluster head updates its cluster with a total rate of $\lambda_c$. In the case of information exchange among the cluster heads, a cluster head updates its neighboring cluster heads as a rate $\lambda_{ca}$ Poisson process and updates its cluster with a total rate of $\lambda_{cb}$, where $\lambda_{ca} + \lambda_{cb} = \lambda$. Thus, when the cluster heads are connected, each cluster head receives source information with a larger rate but updates its cluster with a smaller rate.

The average version age of a subset $S$ that is composed of nodes within a cluster $c$ is still given by (\ref{thm_age_set}) when the cluster heads exchange information in our clustered network topology. This is because when the cluster heads exchange information among themselves, the network topology within a cluster does not change, i.e., the $N_c(S)$ stays the same, and the nodes in a cluster still cannot have a lower version age than their cluster head. 

The only change in (\ref{thm_age_set}) compared to Section~\ref{sect:comm_age} is the average version age of a particular cluster head $c$, $\Delta_c$. As shown in Fig.~\ref{Fig:ring_clusterheads}, even though cluster heads are connected to the nodes in their respective clusters, each cluster head can be updated by the source node or its neighboring cluster heads. That is, to find $\Delta_c$, we only need to look at the first tier of the network, which is the ring gossiping network presented in \cite{Yates21}. Thus, using Lemma~\ref{yates_ring_proof} we find the average version age of a single cluster head when the cluster heads form a ring network as
\begin{align}
    \Delta_c \approx \sqrt{\frac{\pi}{2}}\frac{\lambda_e}{\sqrt{\lambda_s \lambda_{ca}}}\sqrt{m}. \label{Delta_c_ring}
\end{align}
We note that in (\ref{Delta_c_ring}), we have $\sqrt{\lambda_s \lambda_{ca}}$ in the denominator unlike \cite{Yates21} as $\lambda_s$ and $\lambda_{ca}$ are not necessarily equal in our model. We observe in (\ref{Delta_c_ring}) that a single cluster head's average version age approximately scales as $O(\sqrt{m})$ as opposed to $O(m)$ in Theorem~\ref{thm_age_set} since cluster heads now form a ring network. With that, in what follows, we analyze the average version age scaling for different cluster topologies when the cluster heads form a ring network.

\subsection{Version Age in Clustered Disconnected Networks with Connected Cluster Heads}\label{subsect_a}

The network model in this case is as in Section~\ref{Sect:disconn} except that the cluster heads form a ring network in the first hop. Nodes within a cluster are disconnected, i.e., $N_c(S_1) = \emptyset$.
By invoking Theorem~\ref{thm_age_set}, we use the recursion in (\ref{Yates_recursion}) and find the average version age of a single node as
\begin{align}
    \Delta_{S_1} = \Delta_c + k\frac{\lambda_e}{\lambda_{cb}} \approx \sqrt{\frac{\pi}{2}}\frac{\lambda_e}{\sqrt{\lambda_s \lambda_{ca}}}\sqrt{m} + k\frac{\lambda_e}{\lambda_{cb}}, \label{discon_ring_ch}
\end{align}
where the approximation follows from (\ref{Delta_c_ring}).

\begin{theorem}\label{thm_discon_ring_ch}
  In a clustered network of disconnected users when the cluster heads form a ring network, the average version age of a single user scales as $O(n^{\frac{1}{3}})$.
\end{theorem}

Theorem~\ref{thm_discon_ring_ch} follows by selecting, $k=n^{\frac{1}{3}}$ with $m = \frac{n}{k} = n^{\frac{2}{3}}$ in (\ref{discon_ring_ch}) for fixed $\lambda_e$, $\lambda_s$, $\lambda_{ca}$ and $\lambda_{cb}$ that do not depend on $n$. Theorem~\ref{thm_discon_ring_ch} implies that even though nodes in clusters do not gossip, by utilizing the information exchange at the cluster head level, the average version age scaling of an end user can be improved to $O(n^{\frac{1}{3}})$ from $O(\sqrt{n})$ in Theorem~\ref{corr_disconn}. 

Another interesting observation is the parallelism between (\ref{discon_ring_ch}) and (\ref{ring_approximated_age_final}) due to the \emph{almost} additive structure of the average version age in clustered gossip networks. In the case of (\ref{ring_approximated_age_final}), cluster heads are disconnected and nodes in each cluster form a ring network whereas in the case of (\ref{discon_ring_ch}) the network is reversed, i.e., ring network in cluster heads, disconnected network within clusters. Since, we have $n=mk$, both (\ref{discon_ring_ch}) and (\ref{ring_approximated_age_final}) yield the same $O(n^{\frac{1}{3}})$ average version age scaling at the end users indicating that gossiping equally helps improving the average version age scaling at the end users whether it occurs at the cluster head level or within clusters at the end user level even though possibly newer versions of the source information is exchanged at the cluster head level as cluster heads are directly connected to the source node.

\subsection{Version Age in Clustered Ring Networks with Connected Cluster Heads}\label{sect:conn_ch_ring}

Ring networks are formed both at the cluster head level and within clusters at the end user level. By invoking Theorem~\ref{thm_age_set}, we use the recursion in (\ref{Yates_recursion}) and after following similar steps as in Section~\ref{Sect:ring}, we find
\begin{align}
    \Delta_{S_1}\approx   \sqrt{\frac{\pi}{2}}\frac{\lambda_e}{\sqrt{\lambda_s \lambda_{ca}}}\sqrt{m} +  \sqrt{\frac{\pi}{2}}\frac{\lambda_e}{\sqrt{\lambda\lambda_{cb}}}\sqrt{k}. \label{ring_ring_ch}
\end{align}
We note that (\ref{ring_ring_ch}) is the counterpart of (\ref{ring_approximated_age_final}) where the additional ring network topology at the cluster head level is considered.

\begin{theorem}\label{thm_ring_ring_ch}
  In a clustered network with a ring topology in each cluster when the cluster heads form a ring network, the average version age of a single user scales as $O(n^{\frac{1}{4}})$.
\end{theorem}

Theorem~\ref{thm_ring_ring_ch} follows by selecting $m=\sqrt{n}$ with $k = \frac{n}{m} = \sqrt{n}$ in (\ref{ring_ring_ch}) for fixed $\lambda_e$, $\lambda_s$, $\lambda_{ca}$, and $\lambda_{cb}$. Here, we note that, when gossiping is employed both at the cluster head level and at the end user level within clusters through a ring topology, the average version age scaling at the end users is improved from $O(n^{\frac{1}{3}})$ in Theorem~\ref{corr_ring} to $O(n^{\frac{1}{4}})$ in Theorem~\ref{thm_ring_ring_ch}.

Another interesting observation is the fact that since the network topology is symmetric at both levels in this case, and the average version age at the end users is \emph{almost} additive in the average version age at the cluster heads, the average version age scaling optimal $m$ and $k$ values are identical. 

We can also note that the selection of the update rates at the cluster heads, i.e., $\lambda_{ca}$ and $\lambda_{cb}$ in (\ref{ring_ring_ch}) is critical. After the optimal selection of $m=k=\sqrt{n}$, next, we optimize update rates at the cluster heads, i.e., $\lambda_{ca}$ and $\lambda_{cb}$, to further minimize the version age in (\ref{ring_ring_ch}). After selecting $m=k=\sqrt{n}$, $\Delta_{S_1}$ in (\ref{ring_ring_ch}) becomes 
\begin{align}
  \Delta_{S_1}\approx   \sqrt{\frac{\pi}{2}} \lambda_e n^{\frac{1}{4}}\left(\frac{1}{\sqrt{\lambda_s \lambda_{ca}}}+ \frac{1}{\sqrt{\lambda\lambda_{cb}}}\right). \label{ring_ring_ch_v2}  
\end{align}

The minimization of $\Delta_{S_1}$ in (\ref{ring_ring_ch_v2}) is equivalent to solving the following optimization problem
\begin{align}
\label{problem1}
\min_{\{\lambda_{ca},\lambda_{cb} \}}  \quad & \frac{1}{\sqrt{\lambda_s \lambda_{ca}}}+ \frac{1}{\sqrt{\lambda\lambda_{cb}}} \nonumber \\
\mbox{s.t.} \quad & \lambda_{ca}+\lambda_{cb} = \lambda_{c} \nonumber \\
\quad & \lambda_{ca}\geq 0, \quad \lambda_{cb}\geq 0.
\end{align}
We write the Lagrangian for the optimization problem in (\ref{problem1}) as,
\begin{align}
\mathcal{L} =& \frac{1}{\sqrt{\lambda_s \lambda_{ca}}}+ \frac{1}{\sqrt{\lambda\lambda_{cb}}} + \beta (\lambda_{ca}+\lambda_{cb}- \lambda_{c})- \theta_1\lambda_{ca}-\theta_2\lambda_{cb},
\end{align}
where $\theta_1\geq 0$, $\theta_2\geq 0$, and $\beta$ can be anything. We note that the problem in (\ref{problem1}) is jointly convex with respect to $\lambda_{ca}$ and $\lambda_{cb}$. Thus, by analyzing the KKT conditions we can find the optimal solution. We write the KKT conditions as, 
\begin{align}
\frac{\partial \mathcal{L}}{\partial \lambda_{ca}}  =& -\frac{1}{2 \lambda_s} \lambda_{ca}^{-\frac{3}{2}}+\beta -\theta_1 = 0,\label{KKT2_1} \\
\frac{\partial \mathcal{L}}{\partial \lambda_{cb}}  =& -\frac{1}{2 \lambda} \lambda_{cb}^{-\frac{3}{2}}+\beta -\theta_2 = 0.\label{KKT2_2}
\end{align}
Then, by using the KKT conditions in (\ref{KKT2_1}) and (\ref{KKT2_2}), we find the optimal $\lambda_{ca}$ and $\lambda_{cb}$ as
\begin{align}
    \lambda_{ca} &= \frac{\lambda_c \lambda^{\frac{1}{3}}}{\lambda^{\frac{1}{3}}+\lambda_s^{\frac{1}{3}}},\label{soln_lambda_c1_fin}\\
    \lambda_{cb} &= \frac{\lambda_c \lambda_s^{\frac{1}{3}}}{\lambda^{\frac{1}{3}}+\lambda_s^{\frac{1}{3}}}.\label{soln_lambda_c2_fin}
\end{align}

We observe from (\ref{soln_lambda_c1_fin}) and (\ref{soln_lambda_c2_fin}) that when the cluster heads also form a ring network, it is optimal to choose the update rates among the cluster heads, i.e., $\lambda_{ca}$, proportional to the $\frac{1}{3}$-power of the update rate of the end users, i.e., $\lambda^{\frac{1}{3}}$. Similarly, the total update rate allocated by cluster heads to their own clusters should be proportional to the $\frac{1}{3}$-power of the update rate of the source, i.e., $\lambda_s^{\frac{1}{3}}$.\footnote{Similar optimization problems can be formulated for the clustered disconnected networks in Section~\ref{subsect_a}, and for the clustered fully connected networks in Section~\ref{subsect_c}. In order to avoid repetitive arguments, we skip the update rate optimizations of the cluster heads for these parts and provide the analysis only for the clustered ring networks.}

\subsection{Version Age in Clustered Fully Connected Networks with Connected Cluster Heads}\label{subsect_c}

In this case, nodes within a cluster form a fully connected network whereas each cluster head is only connected to its adjacent neighbors in a ring topology. By invoking Theorem~\ref{thm_age_set}, we use the recursion in (\ref{Yates_recursion}) and after following similar steps as in Section~\ref{Sect:fully}, we find
\begin{align}
\Delta_{S_1} \approx \sqrt{\frac{\pi}{2}}\frac{\lambda_e}{\sqrt{\lambda_s \lambda_{ca}}}\sqrt{m} + \frac{\lambda_e}{\lambda}\log k, \label{approximation_ring_ch}
\end{align}
for large $n$, with $\lambda_{cb} = \lambda$. We note that (\ref{approximation_ring_ch}) is analogous to (\ref{approximation}) when the cluster heads form a ring network.

\begin{theorem} \label{thm_fully_ring_ch}
  In a clustered network with a fully connected topology in each cluster when the cluster heads form a ring network, the average version age of a single user scales as $O(\log n)$.
\end{theorem}

Theorem~\ref{thm_fully_ring_ch} follows by noting that since $k=\frac{n}{m}$ we cannot get rid of the $\log n$ term in (\ref{approximation_ring_ch}). Thus, there are multiple $(m,k)$ pairs that result in the same $O(\log n)$ scaling. For example, when $m=1$ and $k = \frac{n}{m}=n$, we obtain $O(\log n)$ scaling. This implies that having a single cluster of $n$ users as in \cite{Yates21}. In addition, selecting $m=\log^2 n$ and $k = \frac{n}{m} = \frac{n}{\log^2 n}$ yields Theorem~\ref{thm_fully_ring_ch} as well, parallel to the discussion after Theorem~\ref{fully_conn_scaling}. 

An interesting observation from Theorem~\ref{thm_fully_ring_ch} is the fact that additional communication at the cluster head level does not improve the average version age scaling at the end users. In a similar fashion, one can show that, even if the cluster heads form a fully connected network among themselves while the nodes in each cluster also form a fully connected network, the average version age at the end users scales as $O(\log n)$. This is due to the fact that the number of clusters $m$ (hence the number of cluster heads) and the number of nodes in each cluster $k$ are such that $n=mk$. Since the level of gossip, i.e., the connectivity among the cluster heads and the nodes in each cluster, cannot be increased beyond fully connected networks, one can conclude that the average version age scaling cannot be improved further than $O(\log n)$ in the considered clustered gossip networks.

We note that once the cluster heads exchange information among themselves, essentially, what we end up with is a hierarchical gossip networks, where in the first level of hierarchy we have $m$ cluster heads, and in the second level of hierarchy we have $mk$ end nodes clustered into $m$ clusters of $k$ nodes each. Inspired by this structure, in the next section, we forego cluster heads and study the version age in hierarchical clustered gossip networks.

\begin{figure}[t]
	\centerline{\includegraphics[width=0.55\columnwidth]{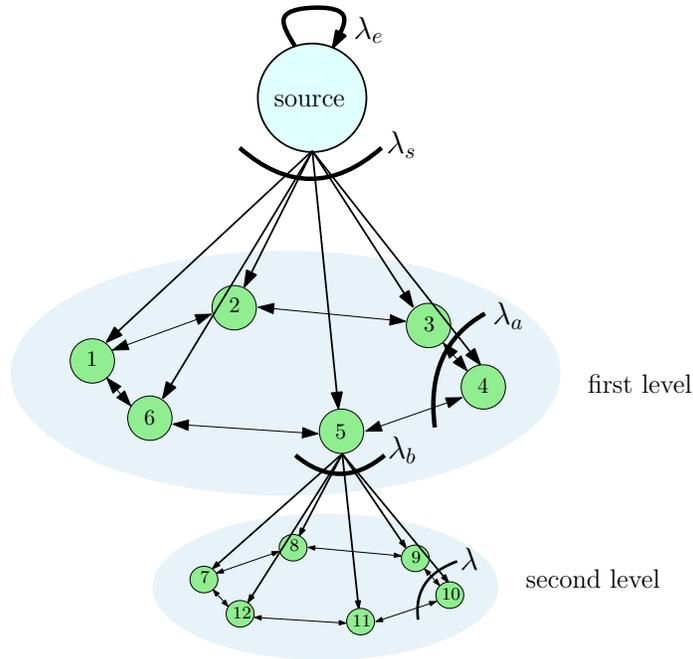}}
	\caption{Two-level hierarchical network model where blue node represents the source and green nodes represent the end users. Here, nodes in each cluster form a bi-directional ring network of $k_1=k_2=6$ nodes. We have a single cluster in the first level and $k_1=6$ clusters in the second level. Here, only one such second level cluster is shown. Other possible network topologies within a cluster are shown in Fig.~\ref{Fig:netw_types}.}
	\label{Fig:hierarchy}
\end{figure}

\section{Version Age in Hierarchical Clustered Gossip Networks}\label{sect:hier}

In this section, we consider a hierarchical clustered gossip network, where at the first level of hierarchy there is a single cluster of $k_1$ nodes. Each node in the first level is directly updated by the source node as a rate $\frac{\lambda_s}{k_1}$ Poisson process. Total update rate of each node is $\lambda$. Each node spends $\lambda_a$ portion of this total rate to update its neighbors within the same cluster at the same hierarchical level and spends $\lambda_b$ portion of the total $\lambda$ rate to update its neighbors in the next hierarchical level such that $\lambda_a+\lambda_b=\lambda$. Each node in the first level is associated with a cluster of $k_2$ nodes in the second level. That is, in the second level, we have $k_2$ clusters and a total of $k_1k_2$ nodes. In Fig.~\ref{Fig:hierarchy}, we show the network model for a two hierarchy levels. We note that at the last level of the network, e.g., the second level in the case of Fig.~\ref{Fig:hierarchy}, nodes use all of their update rate $\lambda$ to update their neighbors within the same cluster. 

Within the scope of this section, we assume that nodes in each cluster at every hierarchical level form a bi-directional ring network.\footnote{One can consider disconnected or fully connected networks within each cluster at each level as well. In our model, since the average version age under fully connected networks cannot be improved beyond $O(\log n)$ as discussed in Section~\ref{sect:conn_ch} and we want to keep our discussion focused, we limit ourselves with a ring network in each cluster at every level.} Let $h$ denote the number of hierarchy levels in the network. Then, at level $i$, with $i \leq h-1$, each node updates each of its two neighbors at level $i$ as a Poisson process with rate $\frac{\lambda_a}{2}$, whereas it updates each of its $k_{i+1}$ child nodes at level $i+1$ as a rate $\frac{\lambda_b}{k_{i+1}}$ Poisson process. We have a total of $\prod_{j=1}^i k_j$ nodes at level $i$ that are grouped into equal-sized clusters of $k_i$ for $i\geq 2$.

Due to the symmetry in the network model, nodes in each hierarchy level $i$ have statistically identical version age processes $\Delta^{i}_{S_1}$. In what follows, we find the average version age expressions of a single node at each hierarchical level.

\begin{theorem}\label{thm_age_hier}
 When the total network of $n$ nodes is grouped into $h$ levels of hierarchical clusters with $k_i$ nodes in each cluster at the $i$th hierarchy level such that $\sum_{i=1}^h\prod_{j=1}^i k_j = n$, the average version age of subset $S$ that is composed of nodes within a cluster $c$ at the $i$th hierarchical level is given by
 \begin{align}
     {\Delta}^{i}_S = \frac{\lambda_e + \lambda_{i-1}(S){\Delta}^{i-1}_{S_1} + \sum_{j \in N_c(S)} \lambda_j(S) {\Delta}^{i}_{S\cup \{j\}} }{\lambda_{i-1}(S) + \sum_{j \in N_c(S)} \lambda_j(S)},\label{Yates_recursion_hier}
 \end{align}
 where $\lambda_{i-1}(S)$ denotes the total update rate at which cluster $c$'s parent node in level $i-1$ updates the nodes in set $S$.
\end{theorem}

The proof of Theorem~\ref{thm_age_hier} follows from that of Theorem~\ref{thm_age_set} by noting that the version age of a node in level $i$ cannot be smaller than that of its parent node in level $i-1$. That is, from the perspective of a node in level $i$, its cluster head in the case of Theorem~\ref{thm_age_set} corresponds to its parent node in level $i-1$.

We first present the results for $h=3$ levels of hierarchy to showcase the version age behavior in the hierarchical gossip networks and then generalize our results to $h$ level of hierarchy.\footnote{For $h=2$ hierarchy levels, the resulting average version age expressions are in the same format as those in Section~\ref{sect:conn_ch_ring} and correspondingly yield an $O(n^{\frac{1}{4}})$ scaling at an end user. This is because the cluster heads forming a ring network at the first tier in Section~\ref{sect:conn_ch_ring} can essentially be thought of as the first level of hierarchy in the context of hierarchical clustered gossip networks analyzed in this section.}

\subsection{Version Age for $h=3$ Hierarchy Levels}

In this case, we have a single cluster of $k_1$ nodes in the first level, $k_1$ clusters of $k_2$ nodes each in the second level, and $k_2$ clusters of $k_3$ nodes in each in the third level of hierarchy so that $n = k_1 + k_1k_2 +k_1k_2k_3$. By using the recursion in Theorem~\ref{thm_age_hier} and noting that at the first level of hierarchy we have the ring network model in \cite{Yates21}, we find
\begin{align}
    \Delta^{1}_{S_1}&\approx   \sqrt{\frac{\pi}{2}}\frac{\lambda_e}{\sqrt{\lambda_s \lambda_a}}\sqrt{k_1}, \label{3hier_eq1} \\
    \Delta^{2}_{S_1}&\approx   \sqrt{\frac{\pi}{2}}\frac{\lambda_e}{\sqrt{\lambda_s \lambda_a}}\sqrt{k_1} +  \sqrt{\frac{\pi}{2}}\frac{\lambda_e}{\sqrt{\lambda_a\lambda_b}}\sqrt{k_2}, \label{3hier_eq2}\\
    \Delta^{3}_{S_1}&\approx   \sqrt{\frac{\pi}{2}}\frac{\lambda_e}{\sqrt{\lambda_s \lambda_a}}\sqrt{k_1} +  \sqrt{\frac{\pi}{2}}\frac{\lambda_e}{\sqrt{\lambda_a\lambda_b}}\sqrt{k_2} +  \sqrt{\frac{\pi}{2}}\frac{\lambda_e}{\sqrt{\lambda\lambda_b}}\sqrt{k_3}. \label{3hier_eq3}
\end{align}

\begin{theorem} \label{thm_3hier}
 In a hierarchical clustered network with $h=3$ hierarchy levels and a ring network in each cluster, the average version age of a single user scales as $O(n^{\frac{1}{6}})$ at every hierarchy level.
\end{theorem}

Theorem~\ref{thm_3hier} follows by observing that $n =k_1 + k_1k_2 +k_1k_2k_3 = k_1(1+k_2(1+k_3)) \approx k_1k_2k_3$ for large $n$. That is, when the number of nodes in the network gets large, order-wise majority of the nodes are located at the final hierarchy level. From the symmetry of the cluster topologies at each hierarchy level and the additive structure observed in (\ref{3hier_eq1})-(\ref{3hier_eq3}), we select $k_1=k_2=k_3=O(n^{\frac{1}{3}})$, which yields an average version age scaling of $O(n^{\frac{1}{6}})$ for all three hierarchical levels.

We note that taking $n \approx k_1k_2k_3$ is as if we assume all the nodes are located at the last hierarchical level of the network so that the $O(n^{\frac{1}{6}})$ scaling we find in Theorem~\ref{thm_3hier} represents a worst case scenario since nodes located at the first two hierarchy levels surely have smaller average version age than the nodes located at the last level of the hierarchy.

Theorem~\ref{thm_3hier} shows that by implementing a three-level hierarchical clustered gossip network structure, we can improve the average version age of a single end-user in our model compared to $O(n^{\frac{1}{2}})$ in \cite{Yates21}, $O(n^{\frac{1}{3}})$ in Section~\ref{Sect:ring}, and $O(n^{\frac{1}{4}})$ in Section~\ref{sect:conn_ch_ring}. The improvement in Section~\ref{Sect:ring} compared to the model in \cite{Yates21} originates from the use of $m$ cluster heads with smaller ring networks under each cluster head compared to the single ring network of $n$ nodes in \cite{Yates21}. When the cluster heads participate in gossip in Section~\ref{sect:conn_ch_ring}, end users have better average version age scaling due to the additional information exchange at the cluster heads. Finally, in this section, through hierarchical placement of clusters, we obtain the same scaling as in Section~\ref{sect:conn_ch_ring} with $h=2$ levels without getting help from any dedicated cluster heads and further improve the scaling result when we employ $h=3$ hierarchy levels. That is, by carefully placing all $n$ nodes into hierarchical clusters of ring networks, we get the best average version age scaling at an end user compared all these network models discussed so far.\footnote{We note that average version age scaling of an end node is improved through hierarchical clustering at the expense of increased number of connections in the network, which may incur additional operational cost to the service provider. This aspect will be discussed in Section~\ref{sect_discuss_conc}.}

\subsection{Version Age for $h>3$ Hierarchy Levels}

In a hierarchical clustered gossip network with $h$ hierarchy levels and a ring network topology in each cluster at every hierarchy level, the average version age of a single node located in hierarchy level $i$ is
\begin{align}\label{hier_age}
    \Delta^{i}_{S_1} = \begin{cases} 
      \sqrt{\frac{\pi}{2}}\frac{\lambda_e}{\sqrt{\lambda_s \lambda_a}}\sqrt{k_i}, & i=1, \\
      \Delta^1_{S_1}  +  \sqrt{\frac{\pi}{2}}\frac{\lambda_e}{\sqrt{\lambda_a\lambda_b}}\sum_{j=2}^i\sqrt{k_j}, & 1<i<h, \\
      \Delta^{i-1}_{S_1} + \sqrt{\frac{\pi}{2}}\frac{\lambda_e}{\sqrt{\lambda\lambda_b}}\sqrt{k_i}, & i=h.
   \end{cases} 
\end{align}

\begin{theorem} \label{thm_Hhier}
 In a hierarchical clustered network with $h$ hierarchy levels and a ring network in each cluster, the average version age of a single user scales as $O(n^{\frac{1}{2h}})$ at every hierarchy level.
\end{theorem}

Theorem~\ref{thm_Hhier} follows by approximating $n \approx \prod_{i=1}^h k_i$ for large $n$ and taking $k_i = n^{\frac{1}{h}}$ for $i \in 1, \ldots, h$. 

\section{Numerical Results}\label{sect:num_res}

\begin{figure}[t]
 	\begin{center}
 	\subfigure[]{%
 	\includegraphics[width=0.49\linewidth]{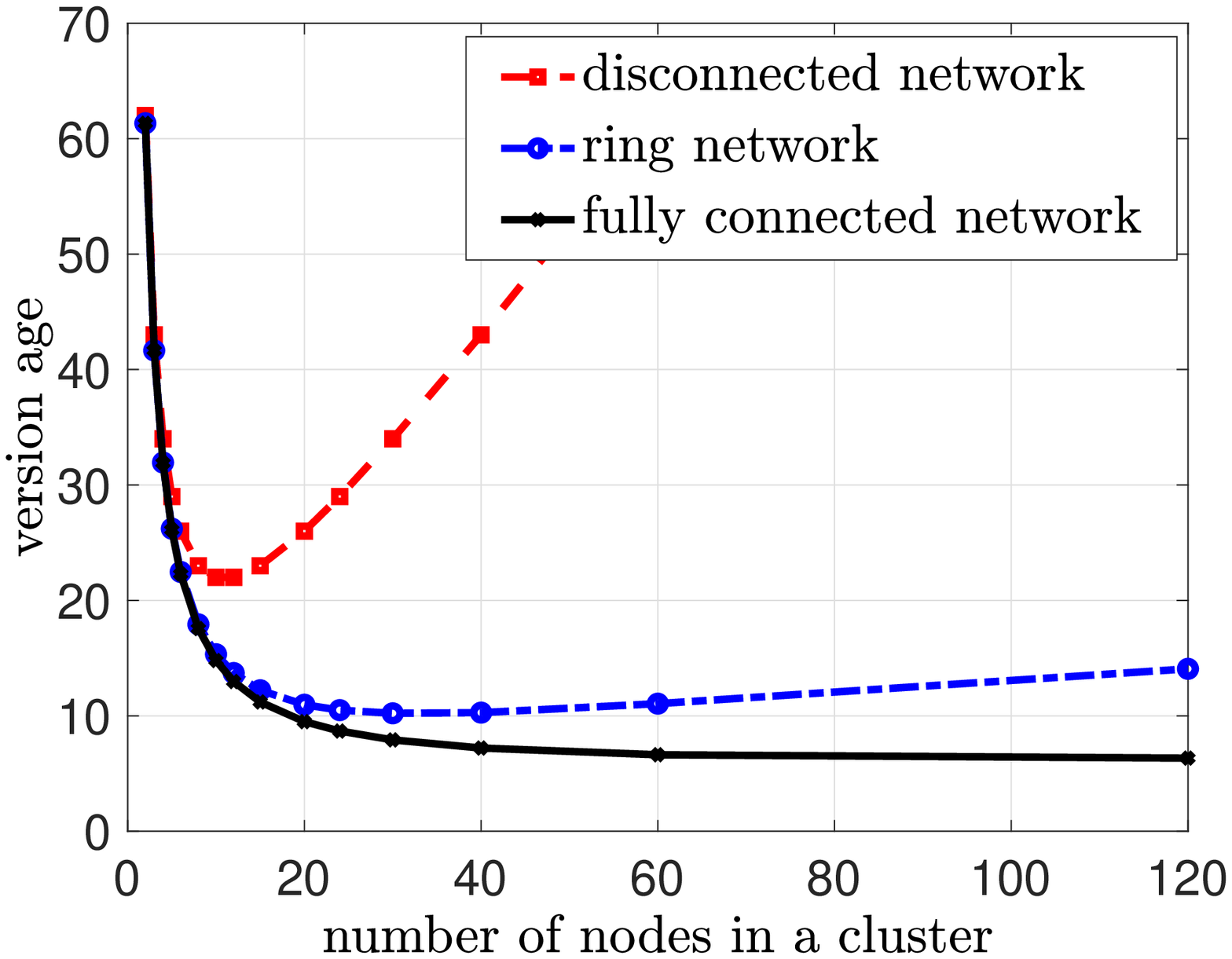}}
 	\subfigure[]{%
 	\includegraphics[width=0.49\linewidth]{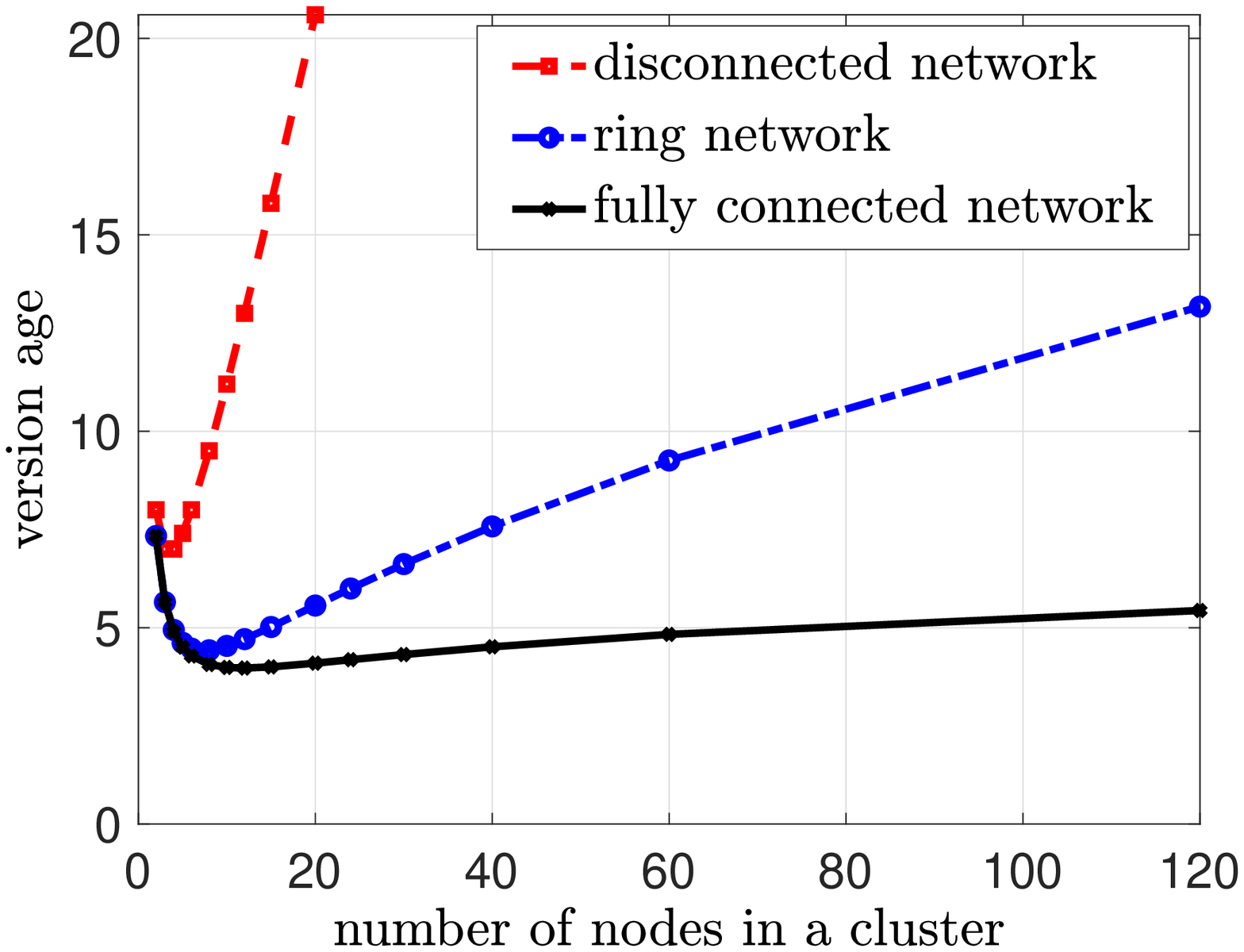}}\\ 
 	\subfigure[]{%
 	\includegraphics[width=0.49\linewidth]{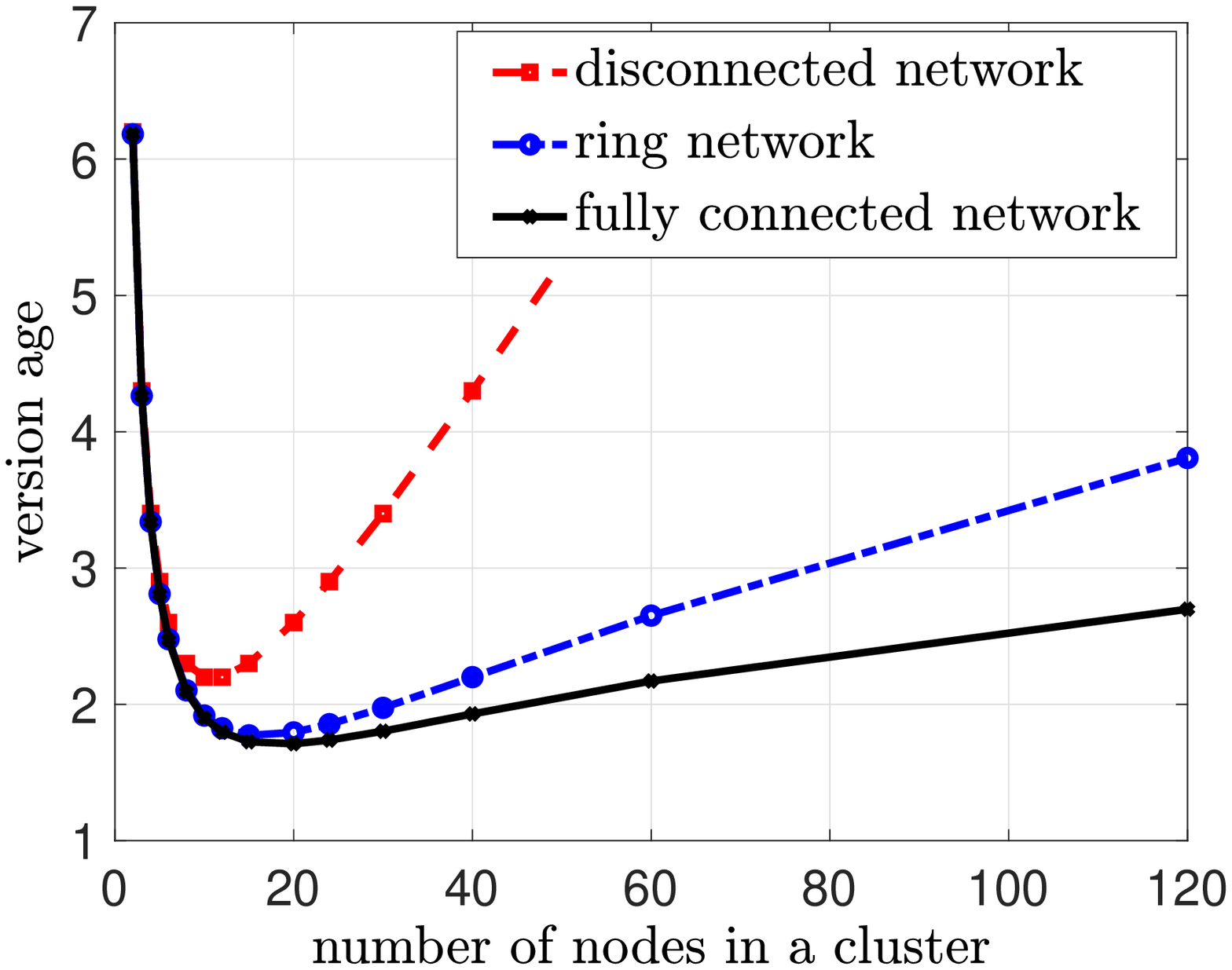}}
 	\subfigure[]{%
 	\includegraphics[width=0.49\linewidth]{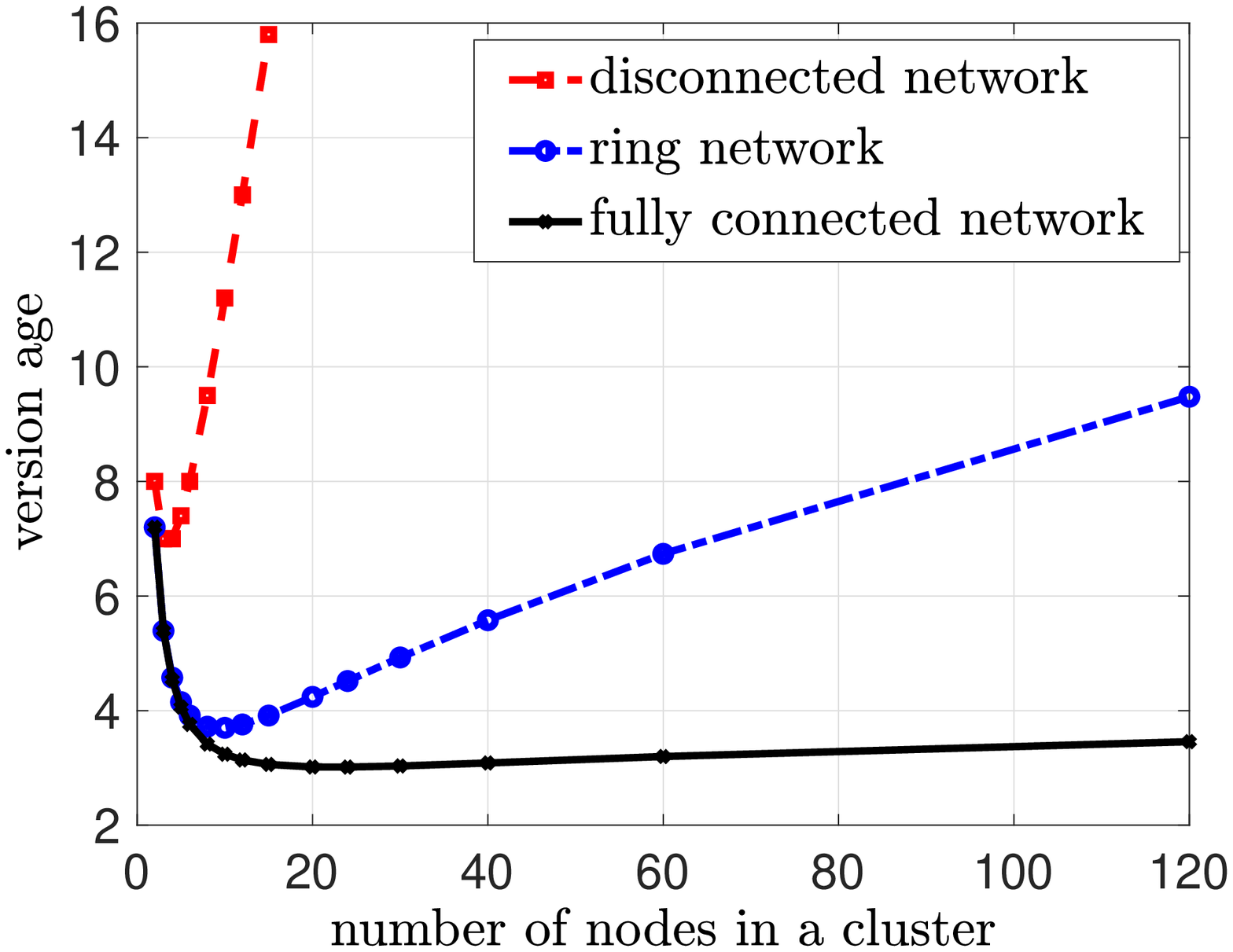}}
 	\end{center}
 	\caption{Version age of a node with fully connected, ring, and disconnected cluster models with $n=120$, (a) $\lambda_e =1$, $\lambda_{s} =1$, $\lambda_{c} =1$, and $\lambda =1$, (b) $\lambda_e =1$, $\lambda_{s} =10$, $\lambda_{c} =1$, and $\lambda =1$, (c) $\lambda_e =1$, $\lambda_{s} =10$, $\lambda_{c} =10$, and $\lambda =1$, (d) $\lambda_e =1$, $\lambda_{s} =10$, $\lambda_{c} =1$, and $\lambda =2$. }
 	\label{Fig:sim_results_all_v2}
\end{figure}

We have seen in Sections~\ref{sect:comm_age}-\ref{sect:hier} that the version age depends on update rates $\lambda_e$, $\lambda_s$, $\lambda_c$, and $\lambda$. In this section, we explore the effects of these rates on the version age via numerical results. In the first four simulations, we consider the model described in Section~\ref{sect:comm_age}.

First, we take $\lambda_e =1$, $\lambda_{s} =1$, $\lambda_{c} =1$, $\lambda =1$, and $n=120$. We plot the version age of a node for the considered cluster models with respect to $k$. We see in Fig.~\ref{Fig:sim_results_all_v2}(a) that for the fully connected cluster model, the version age decreases with $k$ and thus, the version age-optimal cluster size is $k^*=120$, i.e., all $n$ nodes are grouped in a single cluster. In the ring cluster model, the version age is minimized when $k^*=30$. In the disconnected cluster model, the version age is minimized when we have $k^*=10$ (or equivalently $k^*=12$). From these, we deduce that when the topology has less connectivity in a cluster, the optimal cluster size is smaller. Further, a topology with larger connectivity within a cluster achieves a lower version age.

Second, we consider the same setting as in Fig.~\ref{Fig:sim_results_all_v2}(a) but take $\lambda_s =10$ in Fig.~\ref{Fig:sim_results_all_v2}(b). Here, the version age decreases with increasing $k$ at first due to increasing number of connections within a cluster and the increase in the update rate between the source and each cluster head (as the number of clusters decreases with increasing $k$). However, as $k$ continues to increase, the decrease in the update rate from the cluster head to the nodes starts to dominate and the version age increases for all cluster models. In Fig.~\ref{Fig:sim_results_all_v2}(b), we see that the optimal cluster size is $k^*=12$ in fully connected clusters, $k^*=8$ in ring clusters, $k^*=3$ and $k^*=4$ in disconnected clusters.

Third, we increase the update rate of the cluster heads and take $\lambda_{c} =10$. We see in Fig.~\ref{Fig:sim_results_all_v2}(c) that the optimum value of $k$ increases compared to the second case when cluster heads have a larger update rate in all the cluster models. We find $k^*=20$ in fully connected clusters, $k^*=15$ in ring clusters, and $k^*=10$ or $k^*=12$ in disconnected clusters.  

Fourth, we study the effect of update rates among the nodes. For this, we take $\lambda_{c} =10$, $\lambda_e =1$, $\lambda_{s} =1$, $\lambda =2$. We see in Fig.~\ref{Fig:sim_results_all_v2}(d) that as the communication rate between the nodes increases, the optimal cluster size increases, and it is equal to $k^*=24$ in fully connected clusters, and $k^*=10$ in ring clusters. As there is no connection between nodes in the case of disconnected clusters, the optimum cluster size remains the same, i.e., $k^*=3$ or $k^*=4$, compared to Fig.~\ref{Fig:sim_results_all_v2}(b).

\begin{figure}[t]
	\centerline{\includegraphics[width=0.5\columnwidth]{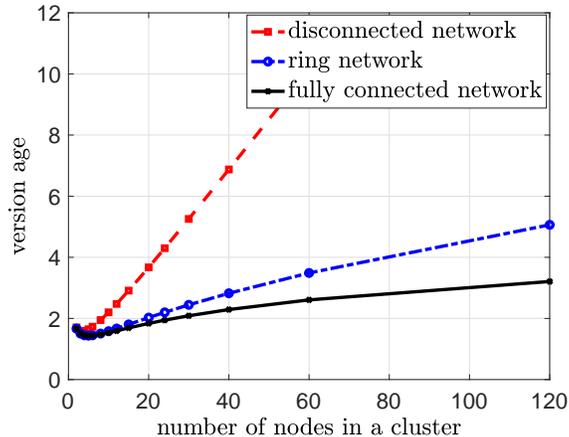}}
	\caption{Version age of a node with fully connected, ring, and disconnected cluster models with $n=120$, $\lambda_e =1$, $\lambda_{s} =10$, $\lambda_{ca} =4$, $\lambda_{cb} =6$, and $\lambda =1$ when cluster heads form a ring network among themselves.}
	\label{Fig:connected_heads}
\end{figure}

\begin{table}[t]
\small
\centering
	\begin{center}
		\begin{tabular}{ | l | l | l |}
			\hline
			& $\text{no gossip at cluster heads}$ & $\text{gossiping cluster heads}$ \\ \hline
			$\text{disconnected}$ & $(2.2000,10)$ & $(1.5936,4)$   \\ \hline
			$\text{ring}$  & $(1.7729,15)$ & $(1.4365,5)$ \\ \hline
			$\text{fully connected}$  & $(1.7111,20)$ & $(1.4291,5)$   \\\hline
		\end{tabular}
	\end{center}
	\caption{Comparison of the $(\Delta^*_{S_1}, k^*)$ pairs with (as in Section~\ref{sect:conn_ch}) and without (as in Section~\ref{sect:comm_age}) gossip at the cluster heads. }
	\label{table:exp_len}
	\vspace{-0.5cm}
\end{table}

Next, we look at the version age when the cluster heads form a ring network as in Section~\ref{sect:conn_ch}. For this simulation, we use the setup of Fig.~\ref{Fig:sim_results_all_v2}(c) and take $\lambda_{ca} = 4$ and $\lambda_{cb}=6$. We see in Fig.~\ref{Fig:connected_heads} for all network types within clusters, i.e., disconnected, ring, and fully connected networks, that version age of a single end node improves when the cluster heads exchange information among themselves. In addition, we observe that with the additional gossip at the cluster heads, the version age optimal cluster size for each case is now smaller. Comparison of the optimal cluster size and the corresponding  minimum version age achieved for each type of network is given in Table~\ref{table:exp_len} with and without gossip at the cluster heads. In Table~\ref{table:exp_len}, we observe that the biggest version age improvement is obtained in the case of disconnected clusters, as the additional communication at the cluster heads is more valuable when nodes within clusters are not connected at all.

In our last numerical result, we look at the version age in hierarchical clustered gossip networks as in Section~\ref{sect:hier} with $h=3$ hierarchy levels. Here, we consider the number of nodes $n=120$, and take $\lambda_e = 1$, $\lambda_s=1$, and total update rate of a node as $\lambda=5$. In this simulation, we consider different $(\lambda_a, \lambda_b)$ pairs, and find the optimal cluster sizes at each hierarchical level $k_1$, $k_2$, and $k_3$ that minimize the version age of a node at the last hierarchical level as these nodes experience the highest version age in the network. We note that the selection of the $(\lambda_a, \lambda_b)$ is important. While choosing a large $\lambda_a$ increases the connectivity between the nodes within clusters, and thus can lower the version age of the nodes at the same hierarchical level, it may also increase the version age of the nodes at the higher hierarchical levels. For this reason, among the $(\lambda_a, \lambda_b)$ pairs given in Table~\ref{table:hierarchy}, we see that choosing $\lambda_a = 2$, and $\lambda_b = 3$ achieves the lowest version age with the optimum cluster sizes equal to $k_1=3$, $k_2= 13$, and $k_3= 2$. We also note that when $\lambda_a$ is relatively small, i.e., $\lambda_a=1$, most of the nodes are placed at the third hierarchical level, i.e., out of $n=120$ nodes, $k_1 k_2 k_3 = 78$ nodes are placed at the third hierarchical level whereas $k_1=3$ nodes and $k_1 k_2 = 39$ nodes are placed in the first and the second hierarchical levels, respectively. As we increase the connectivity among the nodes within the same level ($\lambda_a$), we see that the number of nodes at the upper hierarchical levels increases.                 

\begin{table}[t]
\small
\centering
	\begin{center}
		\begin{tabular}{ | l | l | l | l | l | l |}
			\hline
			$\lambda_a$ & $\lambda_b$ & $k_1$ &$k_2$ &$k_3$ &$\Delta_{S_1}^3$ \\ \hline
			1 & 4 & 3 & 3 & 12 & 3.7992 \\ \hline
			2 & 3 & 3 & 13 & 2 & 3.7239 \\ \hline
			3 & 2 & 3 & 13 & 2 & 3.9143 \\ \hline
			4 & 1 & 8 & 7 & 1 & 4.3354 \\ \hline
		\end{tabular}
	\end{center}
	\caption{Average version age of a single node at the third hierarchy level, $\Delta_{S_1}^3$, with $h=3$, $n=120$, $\lambda_e =1$, $\lambda_{s} =1$, and $\lambda_a+\lambda_b = 5$. For given $(\lambda_a, \lambda_b)$ pairs, we find the optimum $k_1$, $k_2$, and $k_3$ values that minimize $\Delta_{S_1}^3$. }
	\label{table:hierarchy}
	\vspace{-0.5cm}
\end{table}

\section{Discussion \& Conclusion}\label{sect_discuss_conc}

We considered a system where there is a single source and $n$ receiver nodes that are grouped into distinct equal-sized clusters. Nodes in each cluster participate in gossiping to relay their stored versions of the source information to their neighbors. We considered four different types of connectivity among the nodes within the same cluster: disconnected, uni-directional ring, bi-directional ring, and fully connected. First, we considered the use of dedicated cluster heads in each cluster that facilitate communication between the source and the receiver nodes.  For each of these network models, we found the average version age and its scaling as a function of the network size $n$. In particular, we showed that an average version age scaling of $O(\sqrt{n})$, $O(n^{\frac{1}{3}})$, and $O(\log n)$ is achievable per user in disconnected, ring, and fully connected cluster topologies. We then allowed information exchange among the cluster heads and showed that the version age scaling in the case of disconnected and ring networks in each cluster can be improved to $O(n^{\frac{1}{3}})$ and $O(n^{\frac{1}{4}})$, respectively, when the cluster heads also participate in gossiping through a ring formation. Interestingly, we observed that the increased gossip among the cluster heads does not improve the version age scaling when the nodes in each cluster form a fully connected network. Finally, we implemented a hierarchical clustered gossip structure and showed that per user average version scaling of $O(n^{\frac{1}{2h}})$ is achievable in the case of ring networks in each cluster, where $h$ denotes the number of hierarchy levels, even without the aid of the dedicated cluster heads. We numerically determined the optimum cluster sizes that minimize the version age for varying update rates at the source, cluster heads, and the nodes.

Here, the version age scaling improvement from the model in \cite{Yates21} to our hierarchical clustered gossip network design comes at the expense of increased number of connections at the network. For example, considering a ring network in each cluster, the $O(\sqrt{n})$ scaling result of \cite{Yates21} is obtained by $3n$ connections whereas our cluster head-aided $O(n^{\frac{1}{3}})$ scaling result is achieved as a result of a total $3n + n^{\frac{1}{3}}$ connections in the network. When the cluster heads also form a ring network this number increases to $3n + 3\sqrt{n}$ to achieve an $O(n^{\frac{1}{4}})$ scaling result. Finally, in the case of hierarchical clustered gossip networks with $h$ levels, we have a total of $3\sum_{i=1}^h n^{\frac{i}{h}}$ connections in the network which yield a per node average version age scaling of $O(n^{\frac{1}{2h}})$. Considering the operational cost of each such connection, service providers can design the network structure, i.e., the use of cluster heads, number of hierarchy levels along with the level of connectivity in each cluster, based on the operational budget to obtain the desired level of information freshness at the receiver nodes.

As a future direction, one may study the problem where clusters with different network models are present in a system. One may consider the optimization of the update rates from source to the cluster heads connected to the clusters with different network types.    

\bibliographystyle{unsrt}
\bibliography{IEEEabrv,lib_v6,lib}
\end{document}